\documentclass[
    twocolumn,
	aps,
	prd,
	amssymb,
	preprintnumbers,
	secnumarabic,
	nofootinbib]{revtex4-1}
  
\pdfoutput=1

\usepackage{tikz-feynman} 
\usepackage{graphicx}
\usepackage{enumitem}
\usepackage{latexsym}
\usepackage{amsfonts}
\usepackage{amssymb}
\usepackage{color}
\usepackage{amsmath}
\usepackage{slashed}
\usepackage{dcolumn}
\usepackage{verbatim}
\usepackage{float}
\usepackage{multirow}
\usepackage{xspace}
\usepackage[normalem]{ulem}
\usepackage[
pdfauthor={Nirmal Raj}]{hyperref}

\newcommand{\gsim}{\gtrsim}
\newcommand{\lsim}{\lesssim}
\newcommand{\ra}{\rightarrow}

\def\Lc{\mathcal{L}}

\newcommand{\beq}{\begin{equation}}
\newcommand{\eeq}{\end{equation}}
\newcommand{\bea}{\begin{eqnarray}}
\newcommand{\eea}{\end{eqnarray}}
\newcommand{\nn}{\nonumber}

\def\mll{m_{\ell \ell}}

\def\mlq{m_{\acro{LQ}}}
\def\yql{y_{\acro{QL}}}
\def\yul{y_{u\acro{L}}}
\def\yqe{y_{\acro{Q}e}}

\newcommand{\acro}[1]{\textsc{\MakeLowercase{#1}}}

\definecolor{orange}{rgb}{1,0.5,0}
\definecolor{purple}{rgb}{1,0,1}
\definecolor{brown}{rgb}{.7,.2,.2}
\definecolor{violet}{rgb}{.6,.3,.8}

\allowdisplaybreaks

\begin{document}

\title{Hunting leptoquarks in monolepton searches}

\author{Saurabh Bansal}
\author{Rodolfo M. Capdevilla}
\author{Antonio Delgado}
\author{Christopher Kolda}
\author{Adam Martin}
\author{Nirmal Raj}
\affiliation{Department of Physics, University of Notre Dame, 225 Nieuwland Hall, Notre Dame, Indiana 46556, USA}

\begin{abstract}

We show that stringent limits on leptoquarks that couple to first-generation quarks and left-handed electrons or muons can be derived from the spectral shape of the charged-current Drell-Yan process ($p p \ra \ell^\pm \nu$) at Run 2 of the \acro{LHC}. 
We identify and examine all six leptoquark species that can generate such a monolepton signal, including both scalar and vector leptoquarks, and find cases where the leptoquark exchange interferes constructively, destructively or not at all with the Standard Model signal. When combined with the corresponding leptoquark-mediated neutral-current ($p p \ra \ell^+ \ell^-$) process, we find the most stringent limits obtained to date, outperforming bounds from pair production and atomic parity violation.
We show that, with 3000 fb$^{-1}$ of data, combined measurements of the transverse mass in $p p \ra \ell^\pm \nu$ events and invariant mass in $p p \ra \ell^+ \ell^-$ events can probe masses between 8~TeV and 18~TeV, depending on the species of leptoquark, for electroweak-sized couplings. In light of such robust 
sensitivities, we strongly encourage the \acro{LHC} experiments to interpret Drell-Yan (dilepton and monolepton) events in terms of leptoquarks, alongside usual scenarios like $Z'$ bosons and contact interactions.

\end{abstract}
                 
\maketitle

\section{Introduction}

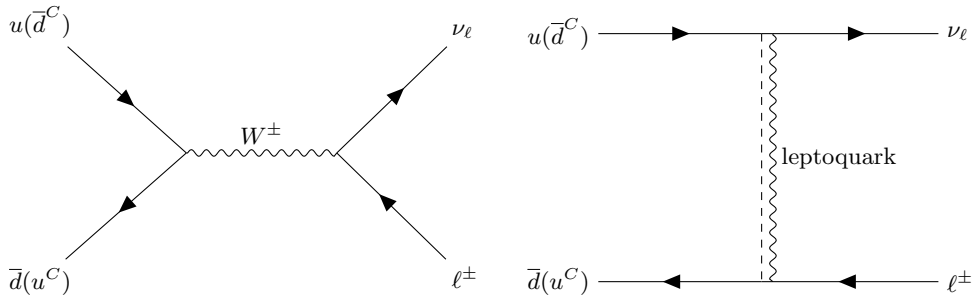
\begin{figure*}[!]	
	\begin{tikzpicture}
	\begin{feynman}
	\vertex (a);
	\vertex [right=2cm of a] (b);
	\vertex [above left=2cm of a] (i1) {$u (\overline{d}^C)$};
	\vertex [below left=2cm of a]  (i2)  {$\overline{d} (u^C)$};
	\vertex [above right=2cm of b] (f1) {$\nu_\ell$};
	\vertex [below right=2cm of b]  (f2) {$\ell^\pm$};
	\diagram* {
		(i1) -- [fermion] (a) -- [fermion] (i2),
		(a) -- [boson, edge label=$W^\pm$] (b),
		(f1) -- [anti fermion] (b) -- [anti fermion] (f2),
	};
	\end{feynman}
	\end{tikzpicture} \quad
\begin{tikzpicture}
\begin{feynman}
\vertex (a) {$u (\overline{d}^C)$};
\vertex [right= 2.7cm of a] (b);
\vertex [right=0.15cm of b](c);
\vertex [right=2.2cm of c]  (f1) {$\nu_\ell$};
\vertex [below=3.3cm of a](a2) {$\overline{d} (u^C)$};
\vertex [right= 2.7cm of a2] (b2);
\vertex [right=0.15cm of b2](c2);
\vertex [right=2.2cm of c2]  (f2) {$\ell^\pm$};
\diagram* {
	(a) -- [fermion] (b) --  (c) -- [fermion] (f1),
	(a2) -- [anti fermion] (b2) --  (c2) -- [anti fermion] (f2),
	(b) -- [scalar] (b2), 	(c) -- [boson, edge label=leptoquark] (c2),
};
\end{feynman}
\end{tikzpicture}
	\caption{
	Feynman diagrams for monolepton production.
    For $S_1$, $S_3$ and $V_2$, the relevant diagrams are those with $\overline{d}^C$ and $u^C$;
	for $U_1$, $U_3$ and $R_2$, the relevant diagrams are those with $u$ and $\overline{d}$.
	Analogous amplitudes exist for dilepton production, $q\overline{q} \ra \ell^+\ell^-$, with $Z$ and $\gamma$ in the $s$-channel of the \acro{SM} diagram.
	These channels may interfere constructively, destructively, or not at all, as explained in the text.
	}
	\label{fig:Feyn}
\end{figure*}

Leptoquarks -- bosonic color triplets with baryon and lepton numbers -- appear in theories of grand unification \cite{Gershtein:1999gp,Dorsner:2005fq}, 
supersymmetry with \acro{R}-parity violation \cite{Barbier:2004ez},
dark matter \cite{1510.03434},
and 
explanations of anomalies in low-energy flavor experiments \cite{1603.04993,1706.07808}.
The \acro{LHC} collaborations hunt them in pair production processes 
\cite{CMS:2016imw,
CMS:2018sgp,
Aaboud:2016qeg},
with current limits having surpassed those placed by the Tevatron and \acro{HERA} 
\cite{GrossoPilcher:1998qt,
Abazov:2008np,
H1:2011qaa,
Abramowicz:2012tg}. 
Leptoquarks coupling to first-generation quarks and electrons or muons are sought in single production processes,
$p p \ra \ell^+ \ell^- j$ \cite{Khachatryan:2015qda}. 
Furthermore, leptoquarks coupling to electrons are strongly constrained by tests of atomic parity violation (\acro{APV}) \cite{1406.4831}.

More recently (see also \cite{Hewett:1997ce,
Rizzo:1997xw,
Choudhury:2002av,
1404.4663} for earlier investigations), it has been argued that leptoquarks can be more effectively sought in dilepton Drell-Yan processes
\cite{
1610.03795}. 
In this paper, we significantly extend that result by showing that 
{\it monolepton}\/ Drell-Yan production at the \acro{LHC},
$p p \ra \ell^\pm \nu$, is actually one of the strongest probes of leptoquarks that couple to left-handed leptons and first generation quarks. Such leptoquarks mediate both monolepton and dilepton production in the $t$- or $u$-channel. Depending on the leptoquark species, the leptoquark contribution may interfere constructively, destructively or not at all with the \acro{SM} Drell-Yan process. 
Each of these cases yields a distinct transverse mass spectrum, whose shape can be used both to rule out the presence of the leptoquark or to differentiate among the species of leptoquarks should evidence for one be found. 
We will show that with the data gathered so far at Run 2 of the \acro{LHC}, monolepton and dilepton production together place the strongest limits on leptoquarks with $O(1)$ couplings and masses $\gsim$ 1 TeV, with monoleptons playing a dominant or significant role in these limits in almost every case considered.

While the precision obtainable in Drell-Yan (dilepton and monolepton) measurements is exploited in this work to constrain leptoquarks, it has also been used to constrain the running of electroweak couplings \cite{Rainwater:2007qa,1410.6810,1602.03877}, 
dark matter \cite{1208.4361,1411.6743,1709.00439},
electroweak precision observables \cite{1609.08157},
and new $Z'$ bosons \cite{1107.5830,1712.02347}. Since several new physics models predict deviations from the \acro{SM} Drell-Yan prediction one may need to use other \acro{LHC} observables or complementary measurements like \acro{APV} to distinguish among those models.

The remainder of this paper is laid out as follows.
In Sec.~\ref{sec:models}, we identify leptoquark species that produce monolepton signals, and spell out coupling structures suited for our study.
In Sec.~\ref{sec:signals}, we describe the unique features of monolepton and dilepton spectra introduced by leptoquark mediation and derive bounds on leptoquarks obtained by comparing to current data.
In Sec.~\ref{sec:concs}, we extend our analysis to the High Luminosity \acro{LHC}, and conclude.

\section{Set-up}
\label{sec:models}

While there are several possible species of leptoquarks distinguished by their spin (0 or 1) and gauge charges \cite{1603.04993}, the species relevant for mediating monolepton production, $u \bar{d} \ra \ell^+ \nu_\ell$, are those that couple to neutrinos (and thus left-handed leptons) and both up- and down-type quarks.
Reference~\cite{1603.04993} catalogs and labels all possible leptoquarks based on their quantum numbers, and from that list one sees immediately that
the above
requirement reduces our options to the
scalars 
$S_1 (\bar{\bf 3}, {\bf 1}, 1/3)$, 
$S_3 (\bar{\bf 3}, {\bf 3}, 1/3)$ and
$R_2 ({\bf 3}, {\bf 2}, 7/6)$, 
and vectors\footnote{In order to realize a fully realistic and unitary model with vector leptoquarks, one must embed them into an appropriate \acro{UV} completion; however, for the purposes of this study, in which the vectors only mediate processes involving light external fermions, such a complete model is not necessary.} 
$U_1 ({\bf 3}, {\bf 1}, 2/3)$,
$U_3 ({\bf 3}, {\bf 3}, 2/3)$ and 
$V_2 (\bar{\bf 3}, {\bf 2}, 5/6)$, 
where the parentheses contain $SU(3)_c \otimes SU(2)_W \otimes U(1)_Y$ quantum numbers.  

Coincidentally, these six leptoquarks span the entire space of leptoquark spins and types of interference in monolepton production. 
More specifically, the three scalar and three vector leptoquarks exhibit destructive interference with the \acro{SM} amplitudes ($S_1, U_1$), constructive interference ($S_3, U_3$), and no interference ($R_2, V_2$). 
Each interference type leads to a distinctive transverse mass spectrum that can be used to place bounds on the leptoquarks and to aid in their differentiation should one be discovered. 

Any leptoquark that can produce a monolepton signature must also be capable of producing a dilepton signature, due to its coupling with the left-handed lepton doublet. We will mention what one can learn about leptoquarks from the dilepton spectrum, though this has been previously studied~\cite{1610.03795} and is not the main purpose of this work. Importantly, unlike the monolepton signal, dileptons can also arise from leptoquark couplings to right-handed quarks or leptons. For this study, we will turn off all couplings that do not generate monoleptons. Turning these couplings back on has no effect on the monolepton signal, but may yield stronger experimental bounds in dilepton channels.
 
For this work, we pick one leptoquark of either spin, the weak singlets $S_1$ and $U_1$, as the prototypes to be discussed in detail. However, results for both monolepton and dilepton signals in all six cases will be shown, as each demonstrates some new features that distinguish it from the others. 

To establish the form of the leptoquark interactions, let us first denote conjugate fermion fields as $\Psi^C \equiv \mathcal{C} \overline{\Psi}^{\rm \acro{T}}$, where $\mathcal{C}$ is the charge conjugation matrix.
The leptoquark $S_1$ generically couples to the diquarks $\overline{Q}^C_LQ_L$ and $\overline{u}^C_Rd_R$ at the renormalizable level, violating baryon number $B$ and leading to dangerously rapid proton decay.
We eliminate these couplings by manually imposing $B$ conservation. 
The interaction Lagrangian for this leptoquark is then given by
\beq
\nn \Lc \supset 
\lambda^{ij}_{\acro{QL}} \overline{Q}^{C a}_{i} \epsilon^{ab} S_1 L^b_{j} + 
\lambda^{ij}_{ue} \overline{u}^C_{i} S_1 e_{j} + {\rm h.c.}~,
\eeq
where the indices $i, j$ run over fermion generations and $a, b$ are $SU(2)$  indices. 
(The above interactions can be found written in terms of 2-component fields and without the conjugate notation in Appendix~\ref{app:twocomponent}.)
In the electroweak broken phase we may expand the above as
\bea
\nn 
\Lc 
&\supset& 
-  (V_{\rm CKM} \ \yql)^{ij} \overline{d}^C_i S_1 P_L \nu_j +
 \yql^{ij} \overline{u}^C_i S_1 P_L e_j \\
&+& y^{ij}_{ue} \overline{u}^C_i S_1 P_R e_j + {\rm h.c.} ~,
\label{eq:LagS1}
\eea
where the transformation $\lambda^{ij} \ra y^{ij}$ accounts for rotations of the fields into their mass basis, and $V_{\rm CKM}$ is the \acro{CKM} matrix\footnote{We take the neutrinos to be massless, and for that reason do not specify the lepton mixing matrix.}.
The coupling $\yql$ leads to a monolepton signal in $u\bar d\ra \ell^+\nu$ and a dilepton signal (both $\propto \yql^2$). 
The $y_{ue}$ coupling contributes to the dilepton signal $u\bar{u} \ra \ell^+\ell^-$ alone, therefore we set it to zero to focus better on the monolepton signature.

We next choose a flavor structure that couples the leptoquark to only the valence quarks in the proton, and to the electron: $y^{ij}_{\acro{QL}}= y^{11}_{\acro{QL}}\delta_{i1}\delta_{j1}$.
Following the terminology of \cite{1610.03795}, we will refer to this scenario as the ``{\tt electroquark}''.
This coupling structure trivially evades flavor constraints \cite{1404.4663} and allows us to focus on the phenomenology of leptoquarks vis-a-vis \acro{LHC} monolepton measurements. 
We could also have chosen to couple to the muon, either in addition to the electron or separately.
The calculations in our analysis would be essentially unchanged, and the bounds obtainable from mono-muon measurements would be similar to those from mono-electron measurements.

The interactions of the $U_1$ leptoquark are given by
\beq
\nn \Lc \supset 
\lambda^{ij}_{\acro{QL}} \overline{Q}_{i} \gamma^\mu U_{1,\mu} L_{j} + 
\lambda^{ij}_{de} \overline{d}_{i} \gamma^\mu U_{1,\mu} e_{j} + {\rm h.c.}~,
\eeq
expanded as
\bea
\nn
\Lc 
&\supset& 
 \yql^{ij} \overline{u}_{i} \gamma^\mu U_{1,\mu} P_L \nu_{j} +
(V_{\rm CKM} \ \yql)^{ij} \overline{d}_{i} \gamma^\mu U_{1,\mu} P_L e_{j} \\
&+& y^{ij}_{de}  \overline{d}_{i} \gamma^\mu U_{1,\mu} P_R e_{j} + {\rm h.c.}~.
\label{eq:LagU1}
\eea
We set $y_{de} = 0$ in order to eliminate extraneous dilepton signals, and choose a flavor structure similar to the $S_1$ interactions: $y^{ij}_{\acro{QL}}= y^{11}_{\acro{QL}}\delta_{i1}\delta_{j1}$.
The couplings of both $S_1$ and $U_1$ (and the rest of the leptoquarks) are assumed real; see Ref.~\cite{1804.01137} for the phenomenology of leptoquark models with \acro{CP}-violation.

The monolepton signal is generated through the Feynman diagrams displayed in Fig.~\ref{fig:Feyn}; analogous diagrams generate the neutral-current process $q\bar{q} \ra \ell^+\ell^-$. The four-fermion operators generated by each species of leptoquark are shown in Table~\ref{tab:monoops}.
One sees, after Fierzing, that the helicity structure of the leptoquark contributions for $S_1$ and $U_1$ matches that of  \acro{SM} $W$-exchange and thus the two will interfere.

In this case, as we explain in Appendix~\ref{app:betas}, the interference in the monolepton channel is destructive, leading to distinctive spectra. For both of these leptoquarks, the dilepton signal also interferes destructively with the \acro{SM}. But unlike the monolepton signal, the sign and size of the interference in the dilepton channel depends on our turning off other couplings that generate purely dilepton signatures.
\begin{table}[t!]
\begin{center}
\begin{tabular}{|c | c |}
\hline
 Leptoquark &  Operator   \\
\hline
\hline 
~~ & ~~ \\[-2mm]
  $S_1$, $S_3$ &  ~~$\yql^2(\overline{u}^C P_L e)(\overline{d}^CP_L \nu)$    \\ [1mm]
 $U_1$, $U_3$ & $\yql^2 (\overline{u} \gamma^\mu P_L \nu)(\overline{d} \gamma_\mu P_L e)$ \\[1mm] 
  $R_2$ &  ~~$\yul\yqe(\overline{u} P_L \nu)(\overline{d} P_R e)$    \\ [1mm]
 $V_2$ & $\yqe\yul(\overline{u}^C \gamma^\mu P_R e)(\overline{d}^C \gamma_\mu P_L \nu)$ \\[1mm] 
 \hline
\end{tabular}
\end{center}
\caption{
Operators responsible for the monolepton signal, 
up to minus signs and factors of \acro{CKM} elements.
The operators for the singlets $S_1$ and $U_1$ and the triplets $S_3$ and $U_3$ involve only left-handed fermions, causing interference with $W$-mediated monolepton production.
Those for $R_2$ and $V_2$ involve right-handed fermions, causing no such interference.
In our analysis of the latter two species, we set $\yul = \yqe$.
}
\label{tab:monoops}
\end{table}

The story with the four remaining leptoquarks ($S_3, R_2, U_3, V_2$) is similar. 
In lieu of writing out their interaction Lagrangians (for which we refer the reader to \cite{1603.04993}), we have simply collected their effective monolepton four-fermion interactions in Table~\ref{tab:monoops}.
For both $S_3$ and $U_3$, the helicity structure of the monolepton amplitude again matches the \acro{SM}; in this case, however, the interference is constructive, which will lead to particularly strong bounds on these leptoquarks. 
Unlike the previous four species of leptoquark, the monolepton signal for the $R_2$ and $V_2$ is generated by the combined action of two different interactions, with couplings $\yul$ and $\yqe$. Furthermore, the helicity structure for the leptoquark-mediated monolepton signal does not match the \acro{SM} (the charged lepton and one of the quarks are right-handed), and hence the contribution of leptoquarks is neither enhanced nor suppressed by interference effects.
For these two species we will set the couplings $\yul = \yqe$ throughout our analysis for simplicity, and denote this common coupling by ``$\yql$" to economize notation.

The dilepton signatures of $S_3, R_2, U_3, V_2$ are more complex than in the $S_1$ or $U_1$ case because the coupling(s) required to generate a monolepton signal generate multiple amplitudes contributing to dileptons, which undergo some combination of constructive, destructive and no interference with corresponding \acro{SM} amplitudes.
We have completed the analysis of these dilepton signals and show the corresponding limits in the following sections. 
Note that our monolepton analysis remains unchanged by the more complicated interference patterns in the dilepton signal, another argument for the importance of a monolepton analysis in searching for, and studying, leptoquarks.

\section{Signals \& Constraints}
\label{sec:signals}

The monolepton signal generated by each species of leptoquark, along with its accompanying dilepton signal, is a function of only two parameters: the mass of the leptoquark, $\mlq$, and the coupling constant, $\yql$. 
As each of the six relevant leptoquarks exhibits a different interference pattern (destructive, constructive or none) and a different spin, all must be considered separately. 
In order to demonstrate clearly the underlying physics, we begin with a parton-level analysis, after which we present the proton-level spectra and constraints from \acro{LHC} measurements. 

\begin{figure}[t!]
\includegraphics[width=0.42\textwidth]{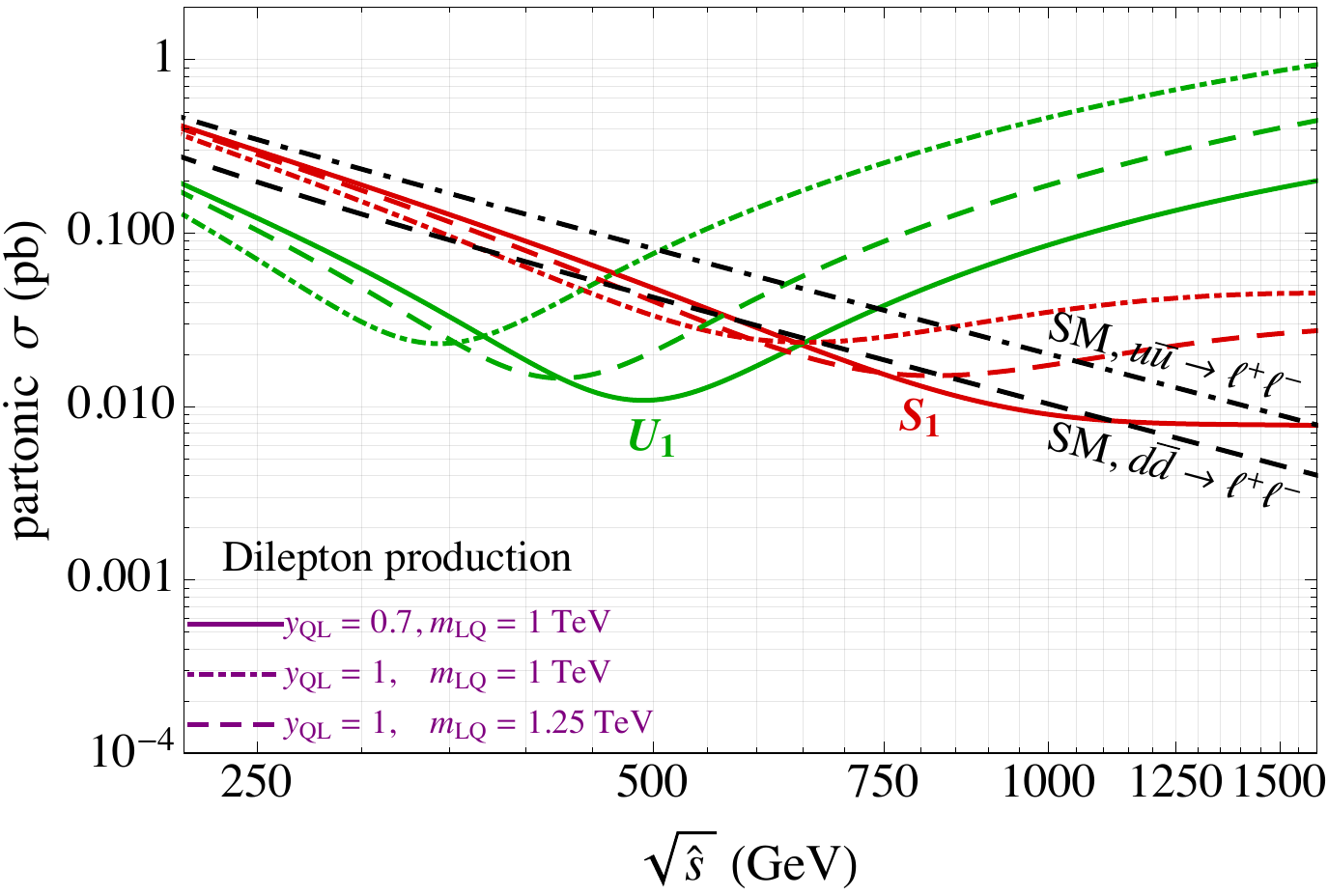} \\
\includegraphics[width=0.42\textwidth]{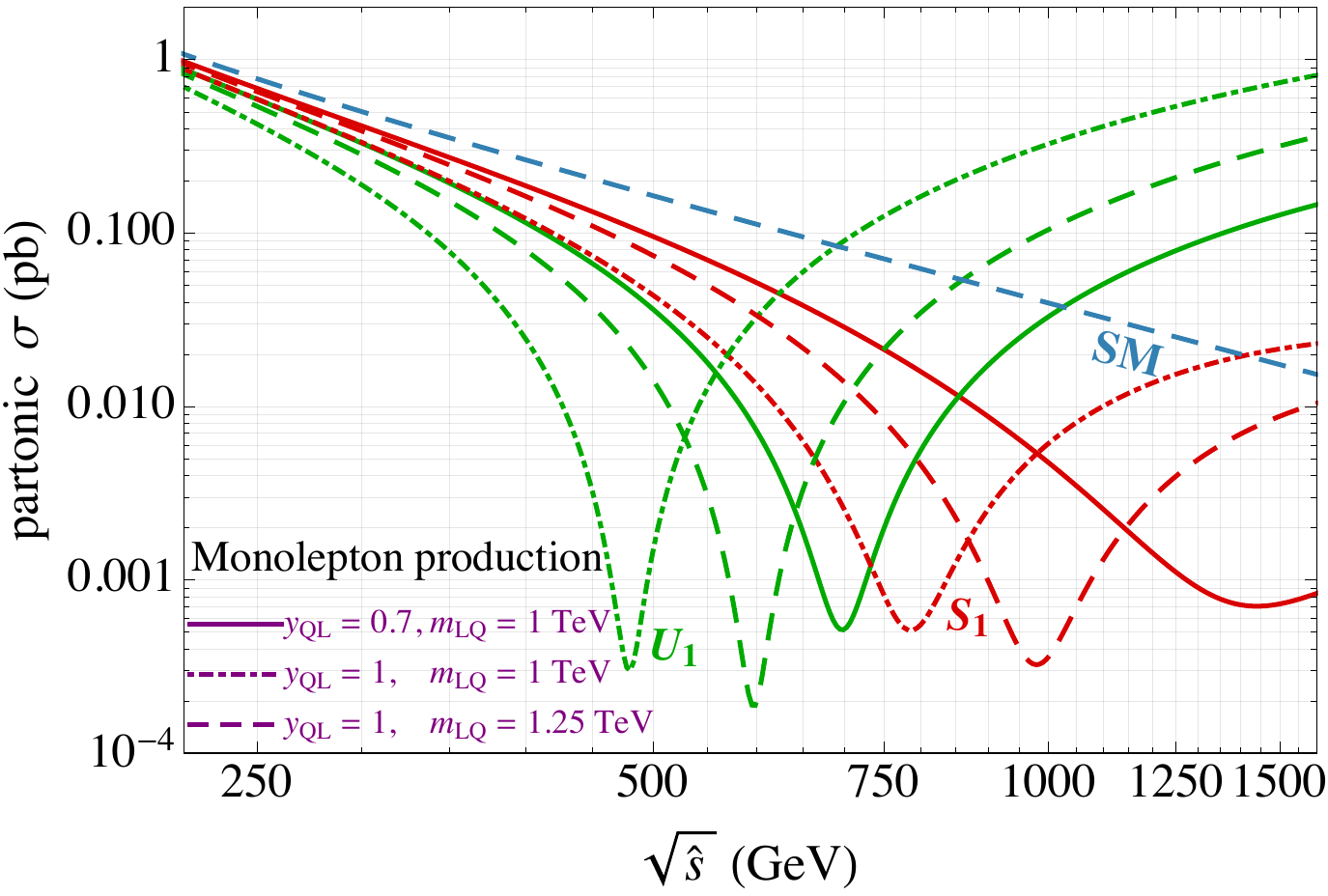} \\
\includegraphics[width=0.42\textwidth]{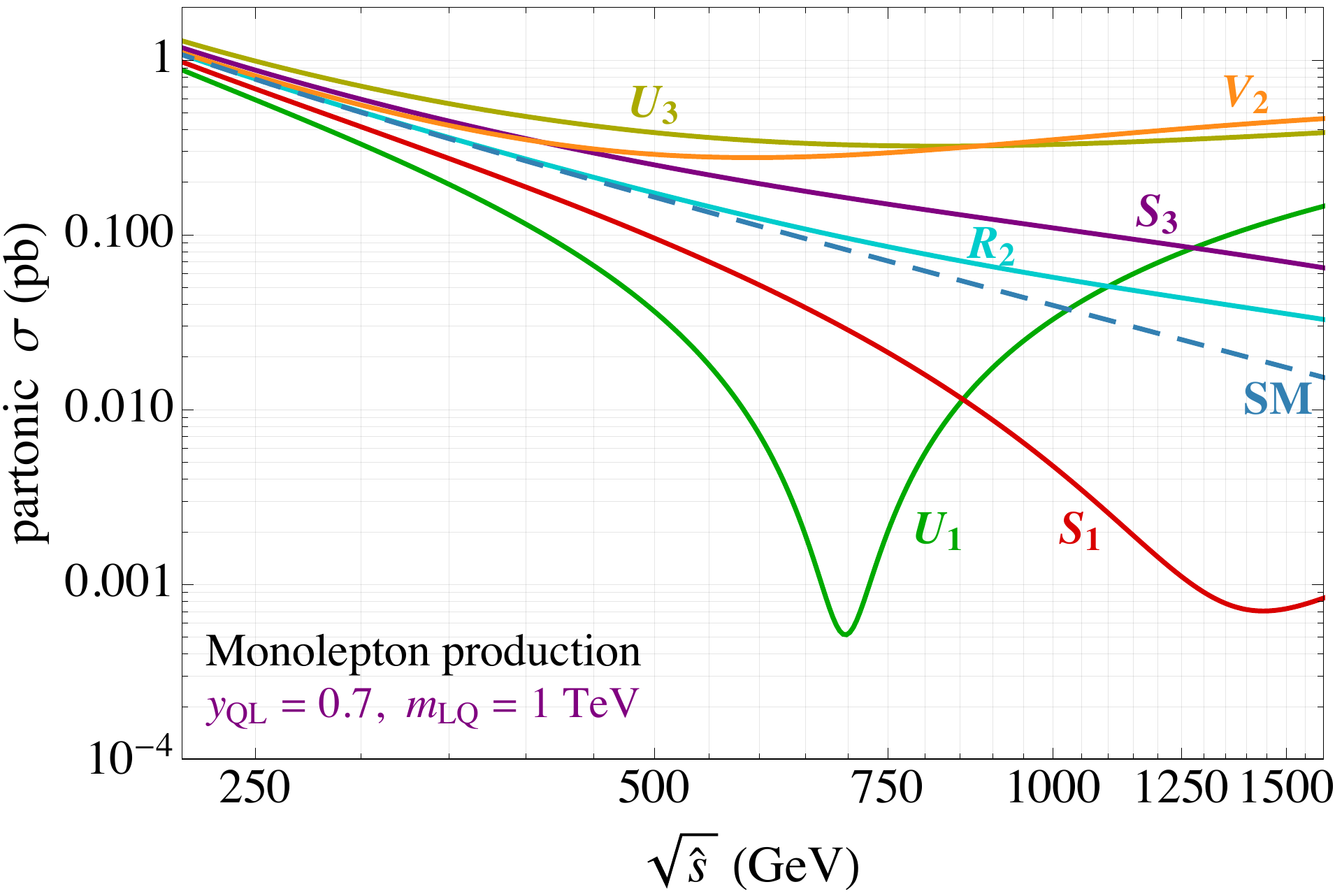} 
\caption{
Partonic cross-sections of monolepton and dilepton production as a function of $\sqrt{\hat{s}}$.
Destructive interferences between \acro{SM} and leptoquark-mediated amplitudes manifest as a significant dip in the mass spectra of these processes.
Constructive or no interferences lead to higher-than-\acro{SM} cross-sections at all $\sqrt{\hat{s}}$.
}
\label{fig:signalsmass}
\end{figure}

In Fig.~\ref{fig:signalsmass} we plot the parton-level cross-sections as a function of the invariant mass, $\sqrt{\hat{s}}$, for both dileptons (top panel) and monoleptons (middle and bottom panels). 
In the first two panels, we restrict ourselves to the spin-0 $S_1$ and spin-1 $U_1$ leptoquarks, but
depict variations of the signal with changing leptoquark couplings and masses. 
To that end, we have chosen three benchmarks: $(\yql, \mlq$/TeV) = $(0.7, 1)$, $(1, 1)$, $(1, 1.25)$. 
(We choose these relatively light leptoquark masses for the figures in this section in order to clearly demonstrate to the eye the effects of the new physics, and especially the effects of interference, on the mono- and dilepton signals. As we will see, current data will push the bounds on the leptoquarks significantly above the 1~TeV mass range.)
One notices in the cross sections unmistakeable dips in both the monolepton and dilepton cross-sections, coming from destructive interference between the \acro{SM} and leptoquark amplitudes. 
Because of the interference, the dips in the cross-section do not occur at $\mlq$ but at energies that depend on both the mass and coupling, $\yql$. 
As expected, the dip moves to larger $\sqrt{\hat s}$ with either increasing $\mlq$ or decreasing $\yql$. 
It is also noteworthy that the dip occurs at lower $\sqrt{\hat s}$ for the vector $U_1$ than for the scalar $S_1$. 
If we were to repeat Fig.~\ref{fig:signalsmass} for a leptoquark with positive interference, such as $U_3$, the result would not be the mirror image of $U_1$. Instead, for this benchmark coupling, one would see a rapid and featureless rise above the $\acro{SM}$, as both the interference piece and the new physics-squared piece of the cross-section have the same sign. 

At values of $\sqrt{\hat{s}}$ larger than the location of the dip and for the value of coupling of $0.7$, we notice an increase in cross-sections for both $S_1$ and $U_1$, where the contribution to the cross-section from the new physics amplitude-squared dominates over the interference term. 
This increase is much steeper for the spin-1 $U_1$ than for the spin-0 $S_1$ because
in the limit $\sqrt{\hat{s}} \gg \mlq$, the cross-section near $\cos\theta^* = 1$ diverges, where $\cos\theta^*$ is the (center-of-momentum frame) scattering angle.
In practice, this divergence of the cross-section near $\cos\theta^* = 1$ is mitigated by kinematic cuts applied for lepton reconstruction.
This phenomenon is kinematically analogous to Rutherford scattering (which proceeds via photon exchange), where the cross-section becomes infinite in the forward ($\cos\theta^* \ra 1$) direction. 
The effect is present for all vector leptoquarks and, 
as we will see later, plays a role in the sensitivity of Drell-Yan processes to vector leptoquarks.

The bottom panel of Fig.~\ref{fig:signalsmass} shows the parton-level monolepton cross section for each of the six leptoquarks, taking $(\yql, \mlq$/TeV) = $(0.7, 1)$. 
Here one can easily see the effects of interference (or lack thereof) on the shape of the spectrum, especially again in the monolepton channel: destructive interference in $S_1$ and $U_1$, constructive in $S_3$ and $U_3$, and none in $R_2$ and $V_2$. 
(The signs of the interference terms are discussed in Appendix~\ref{app:betas}.)
While it is impossible to tell from these parton-level figures the relative strength of the bounds implied by the different interference patterns, it should be obvious that the shapes provide a crucial way for differentiating among the different species of leptoquarks, should a signal be observed.\\

\begin{figure*}[t!]
\includegraphics[width=0.45\textwidth]{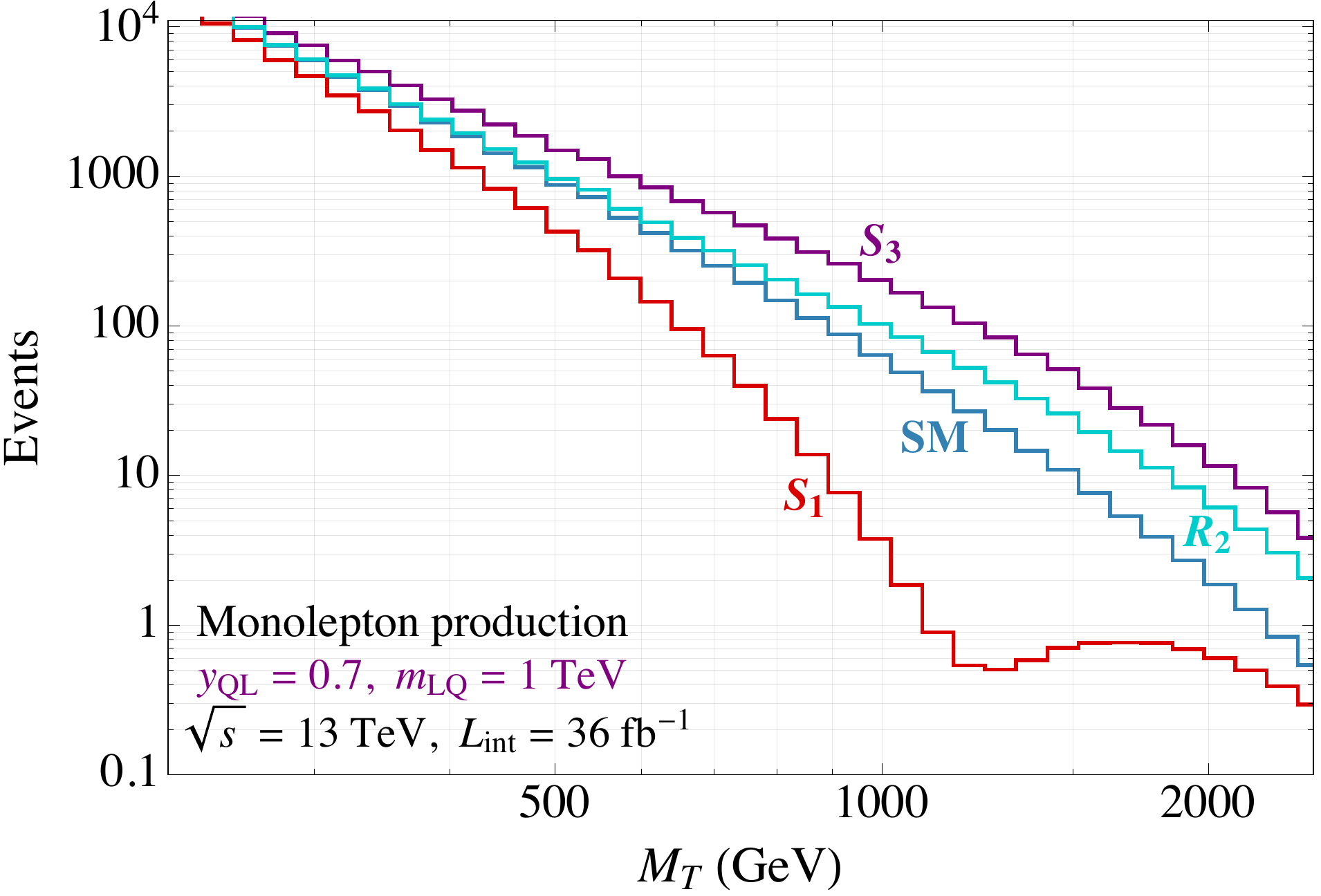} \quad\quad
\includegraphics[width=0.45\textwidth]{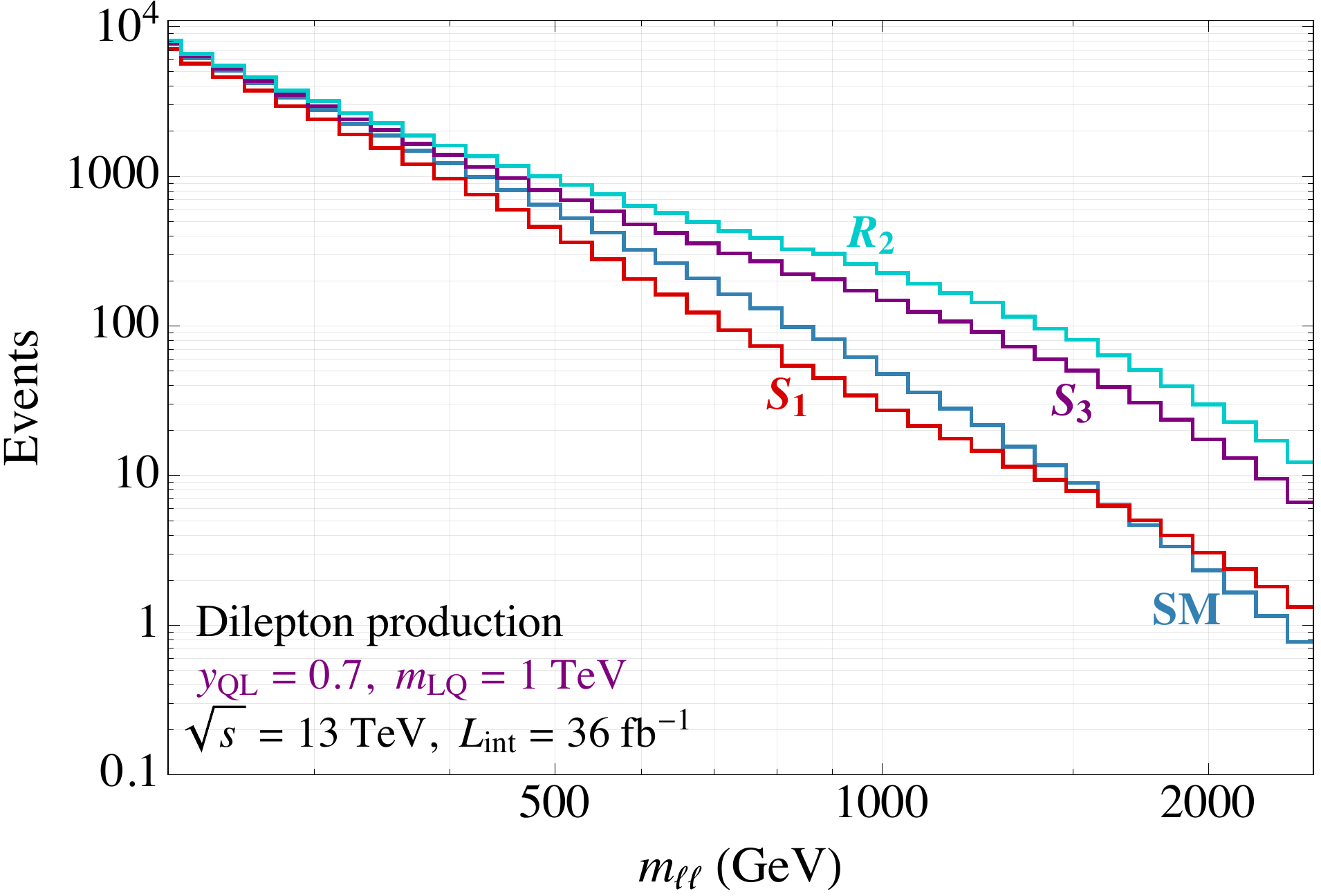} \\
\includegraphics[width=0.45\textwidth]{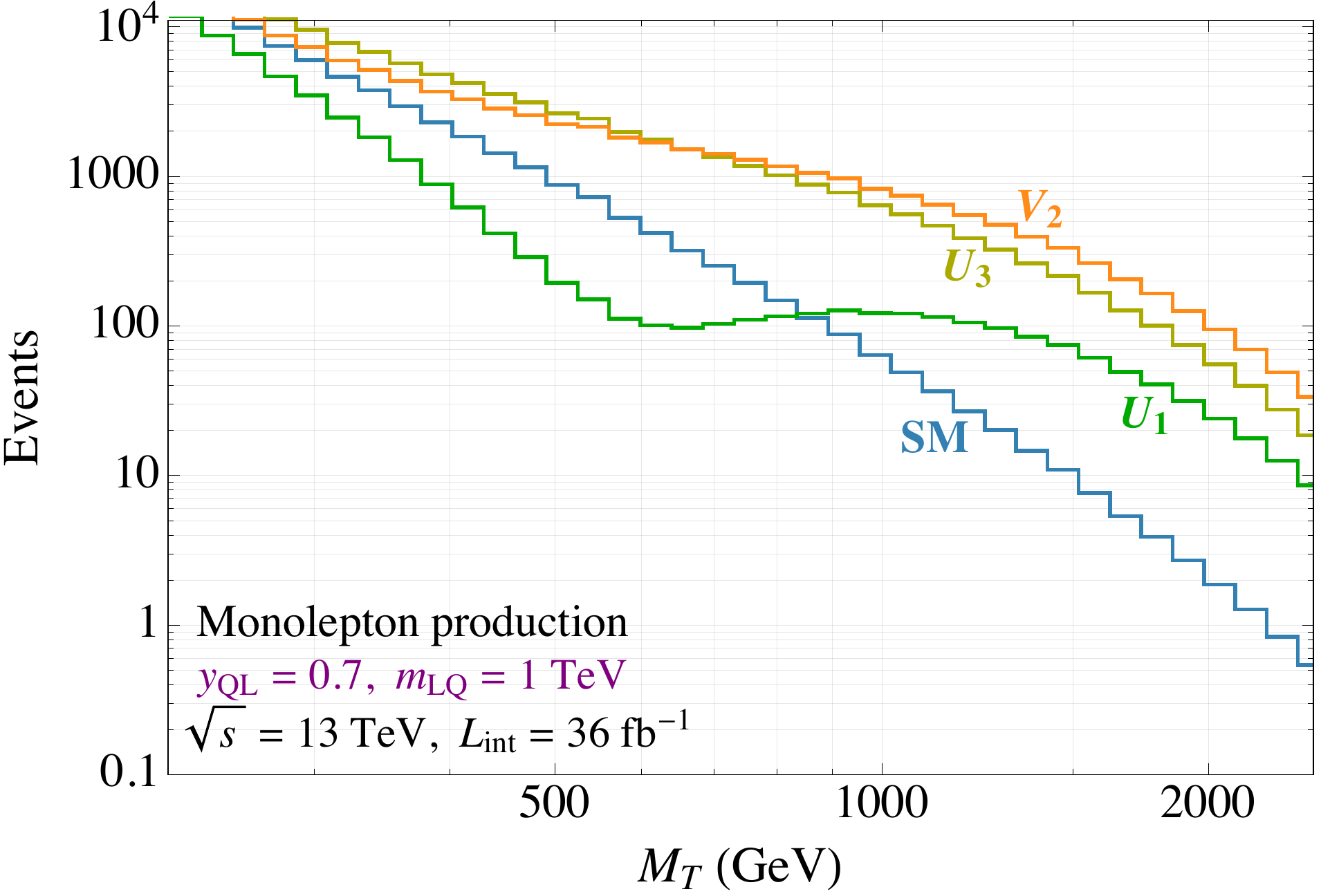} \quad\quad
\includegraphics[width=0.45\textwidth]{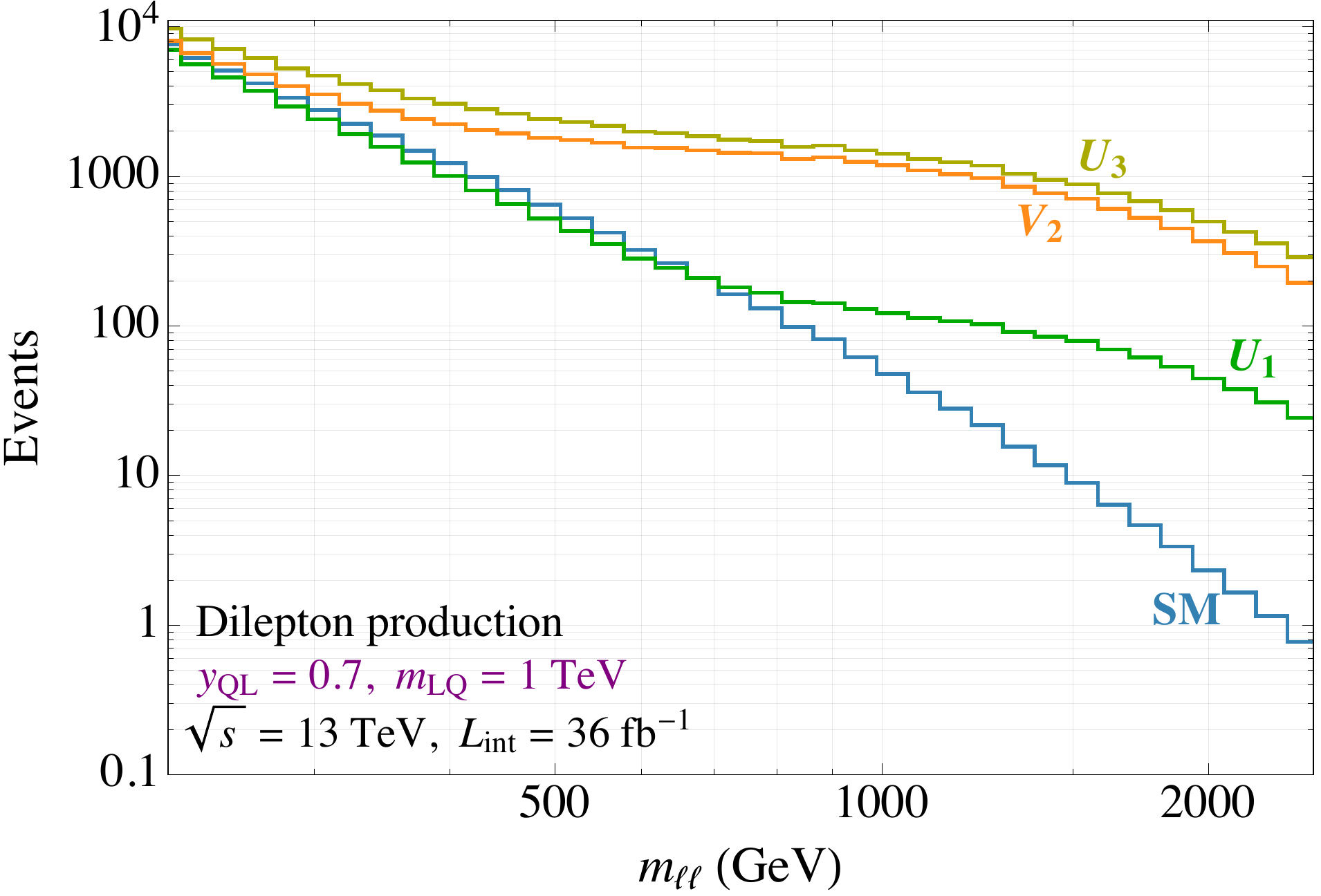} \\
\caption{
Proton-level transverse mass (invariant mass) distributions of monolepton (dilepton) production 
for $\yql = 0.7$ and $\mlq = 1$ TeV.
The upper (lower) panels depict signals of the spin-0 (spin-1) leptoquarks.
}
\label{fig:signalsmassproto}
\end{figure*}

\begin{figure*}[t!]
\begin{center}
\includegraphics[width=0.45\textwidth]{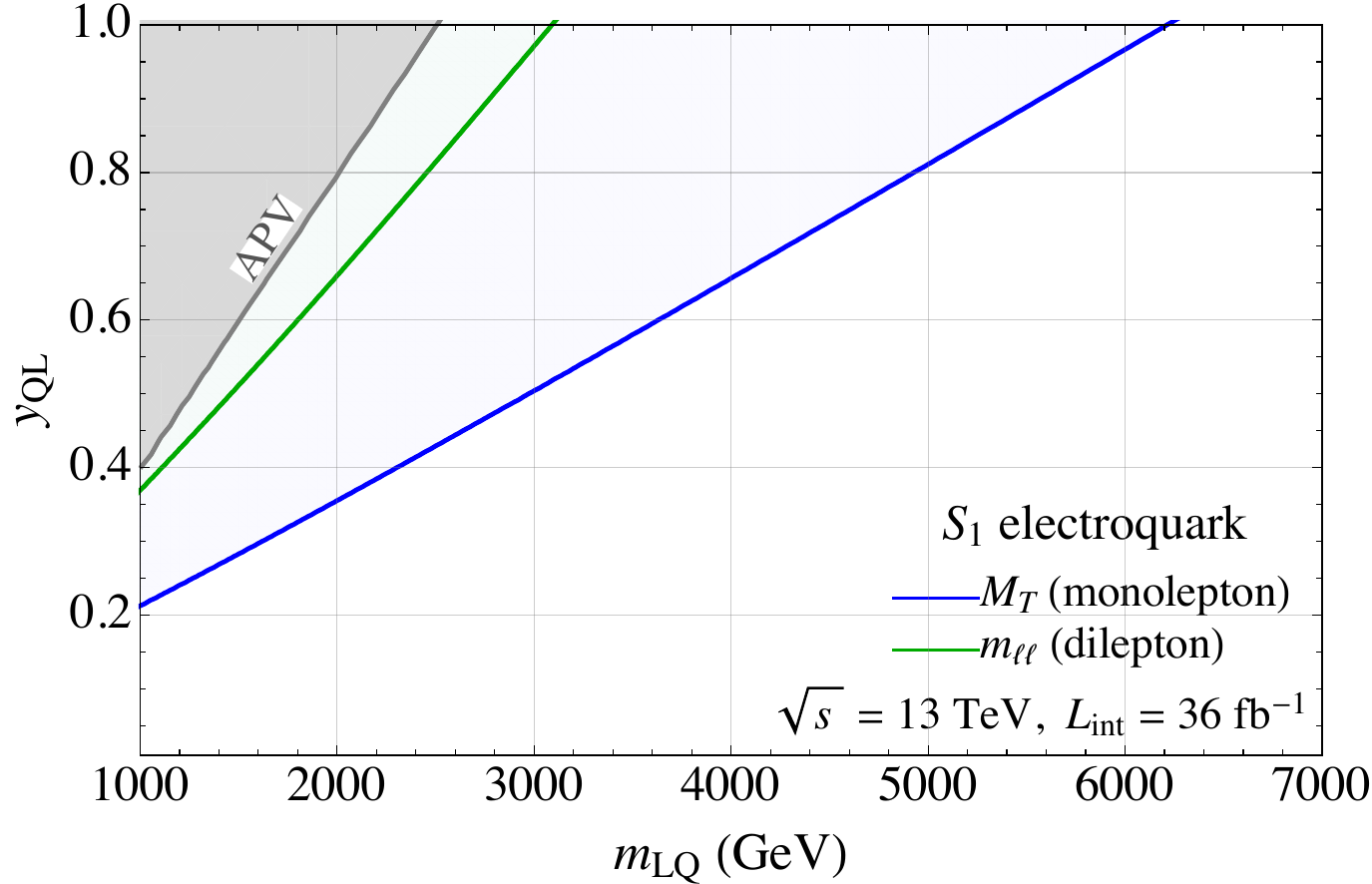} \quad\quad
\includegraphics[width=0.45\textwidth]{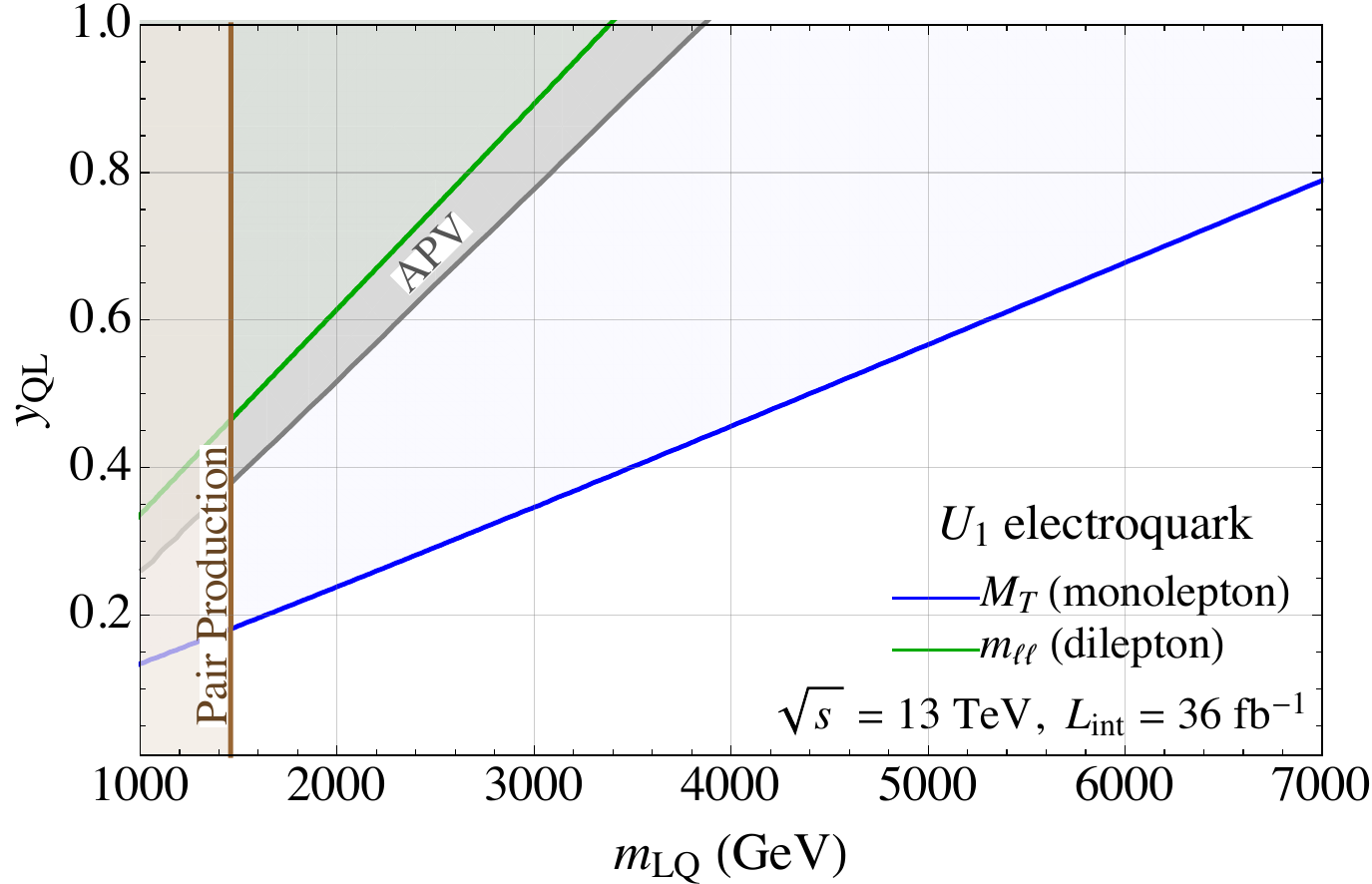} \\
\includegraphics[width=0.45\textwidth]{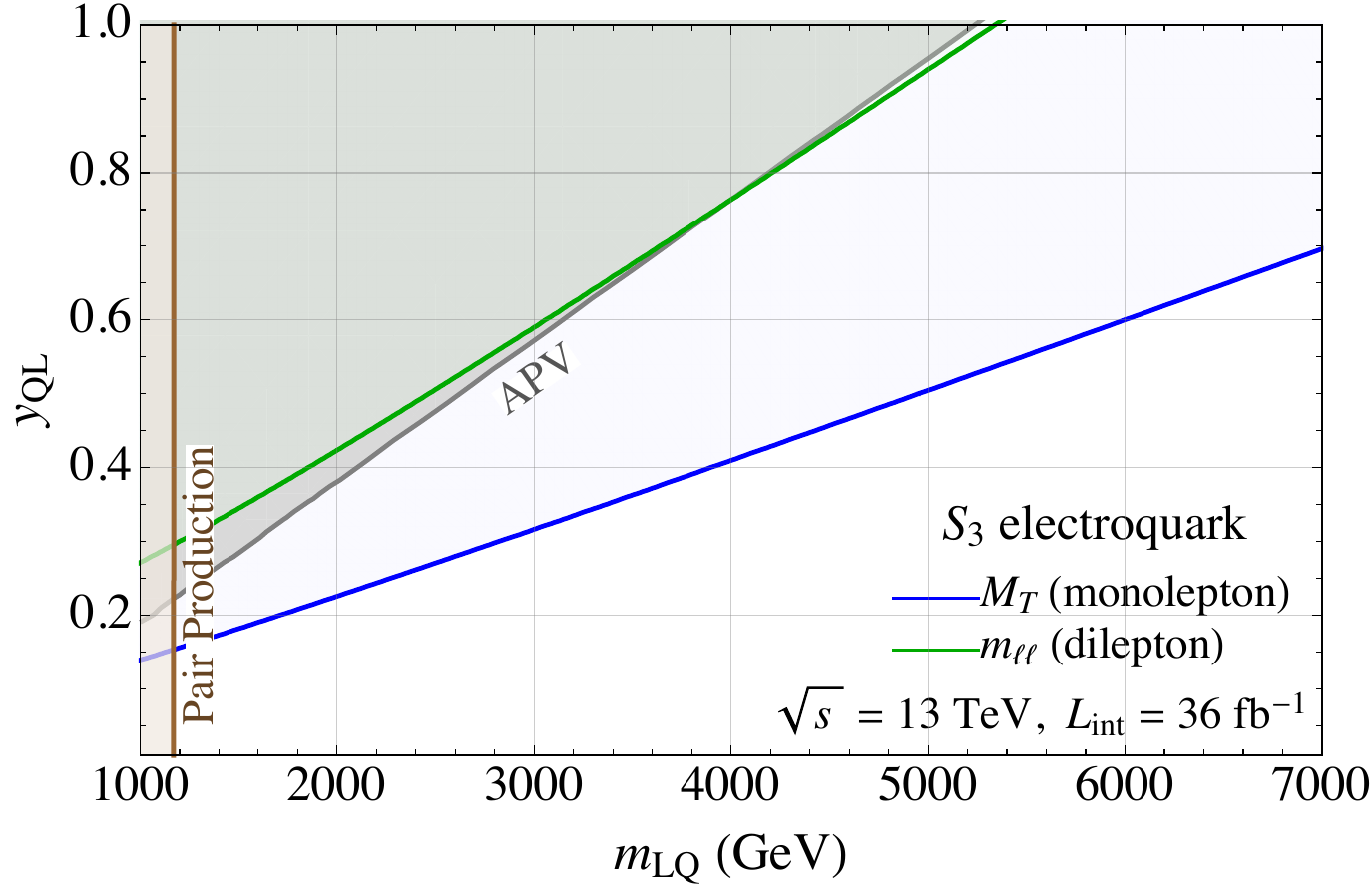} \quad\quad
\includegraphics[width=0.45\textwidth]{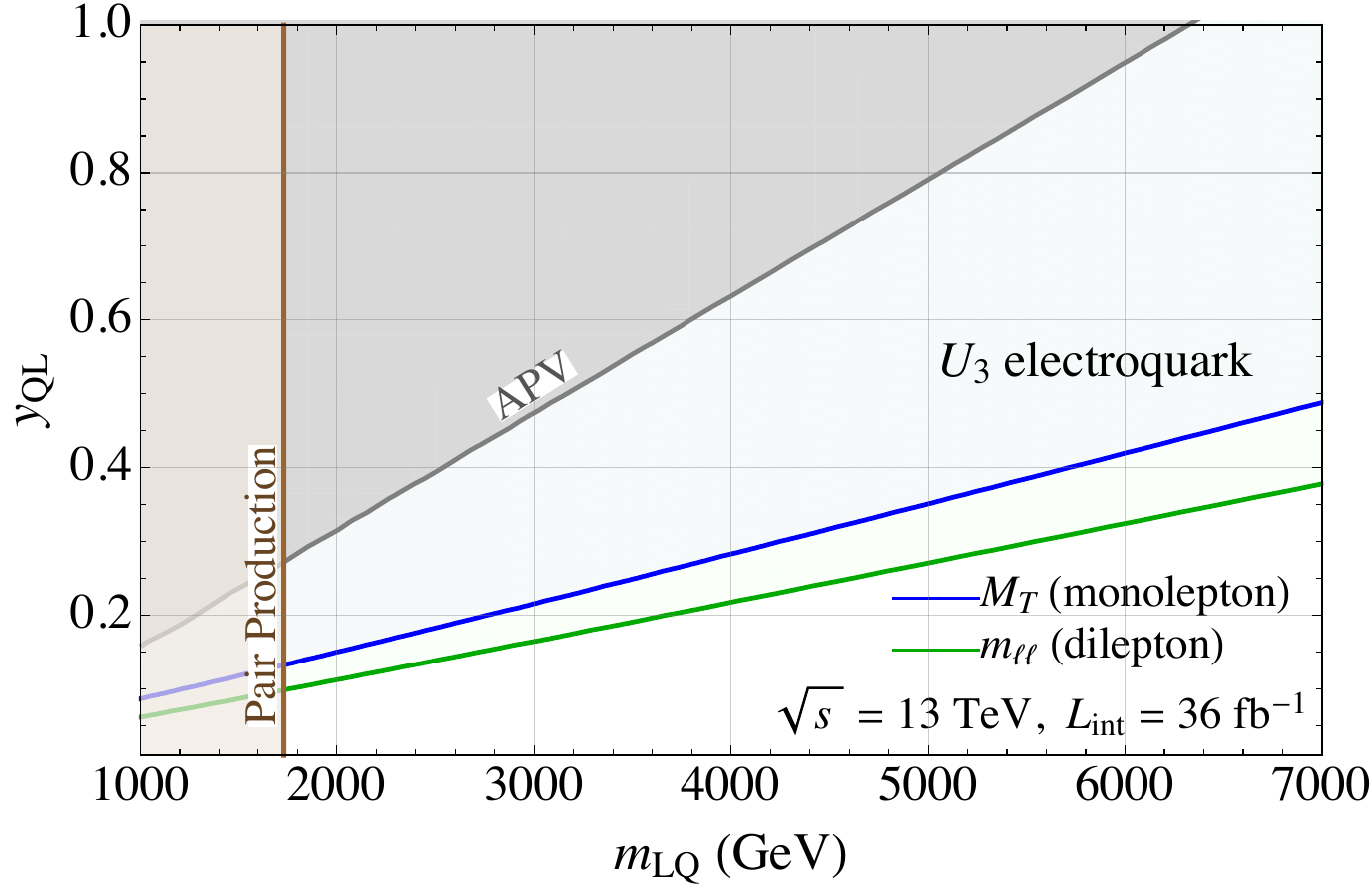} \\
\includegraphics[width=0.45\textwidth]{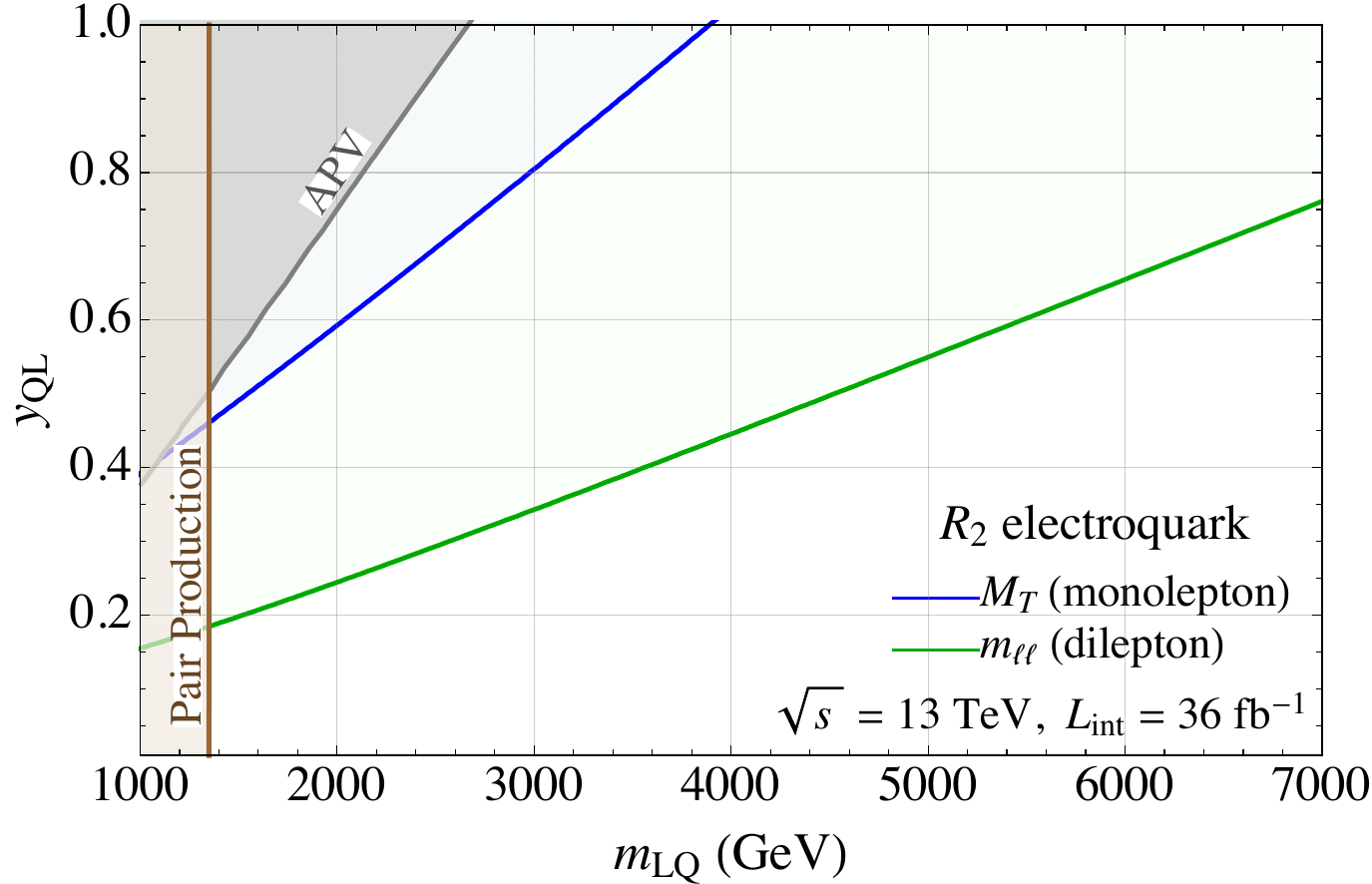} \quad\quad
\includegraphics[width=0.45\textwidth]{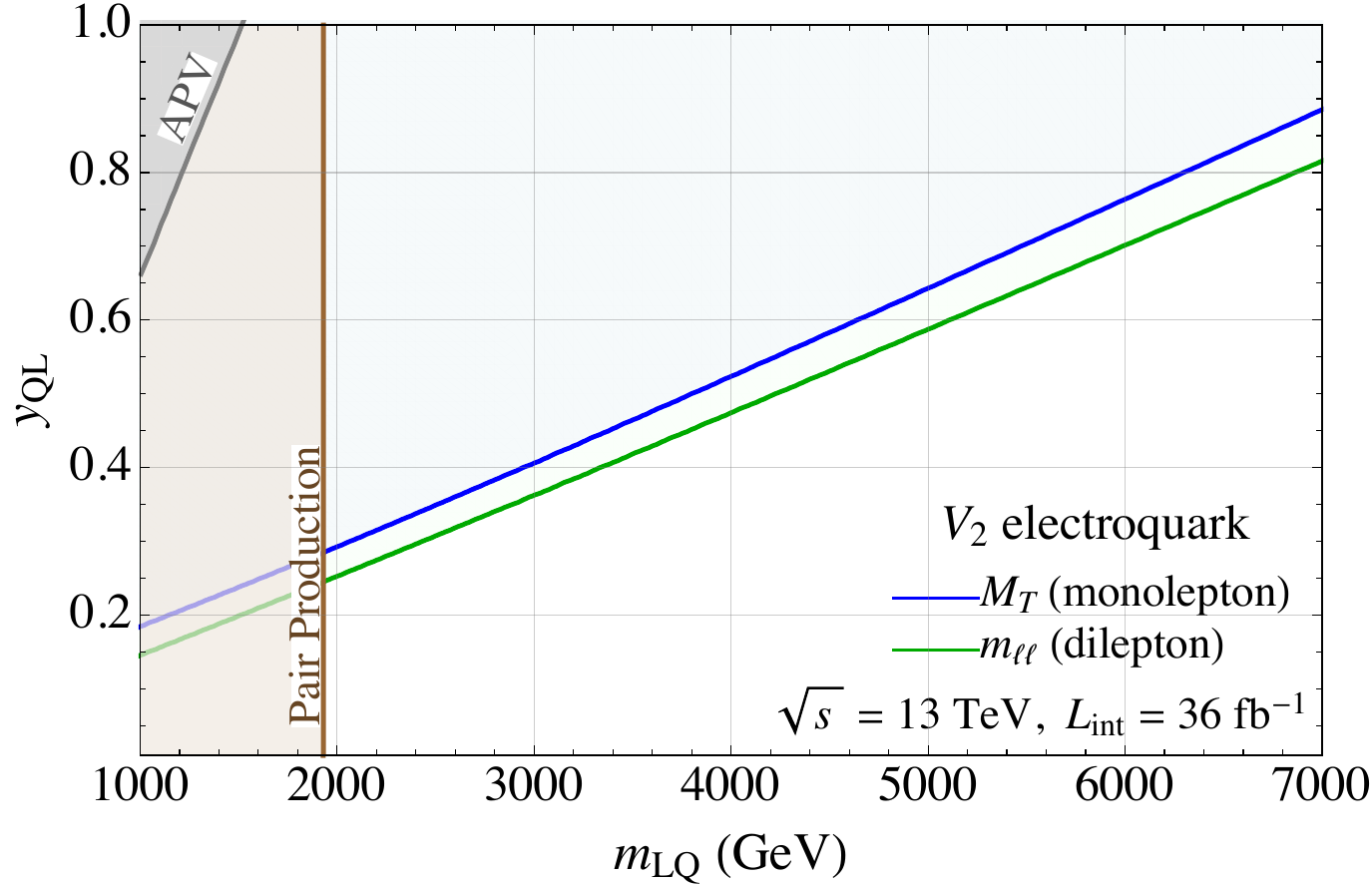} \\
\end{center}
\caption{
The 95\% \acro{C.L.} limits at the 13 TeV \acro{LHC} with 36 fb$^{-1}$ of data, 
on the $S_1, S_3, R_2$ (spin-0) and $U_1, U_3, V_2$ (spin-1) leptoquarks in our set-up, using reported measurements of the monolepton transverse mass and dilepton invariant mass distributions \cite{1706.04786,1707.02424}.
Regions above the corresponding curves are excluded.
Also shown are the 95\% \acro{C.L.} limit from direct pair-production searches \cite{1706.05033,CMS:2016imw}
(brown-shaded regions excluded) 
and the 2$\sigma$ limit on {\tt  electroquarks} from measurements of atomic parity violation 
(grey-shaded regions excluded).
}
\label{fig:limits}
\end{figure*}

\begin{figure}[t!]
\begin{center}
\includegraphics[width=0.45\textwidth]{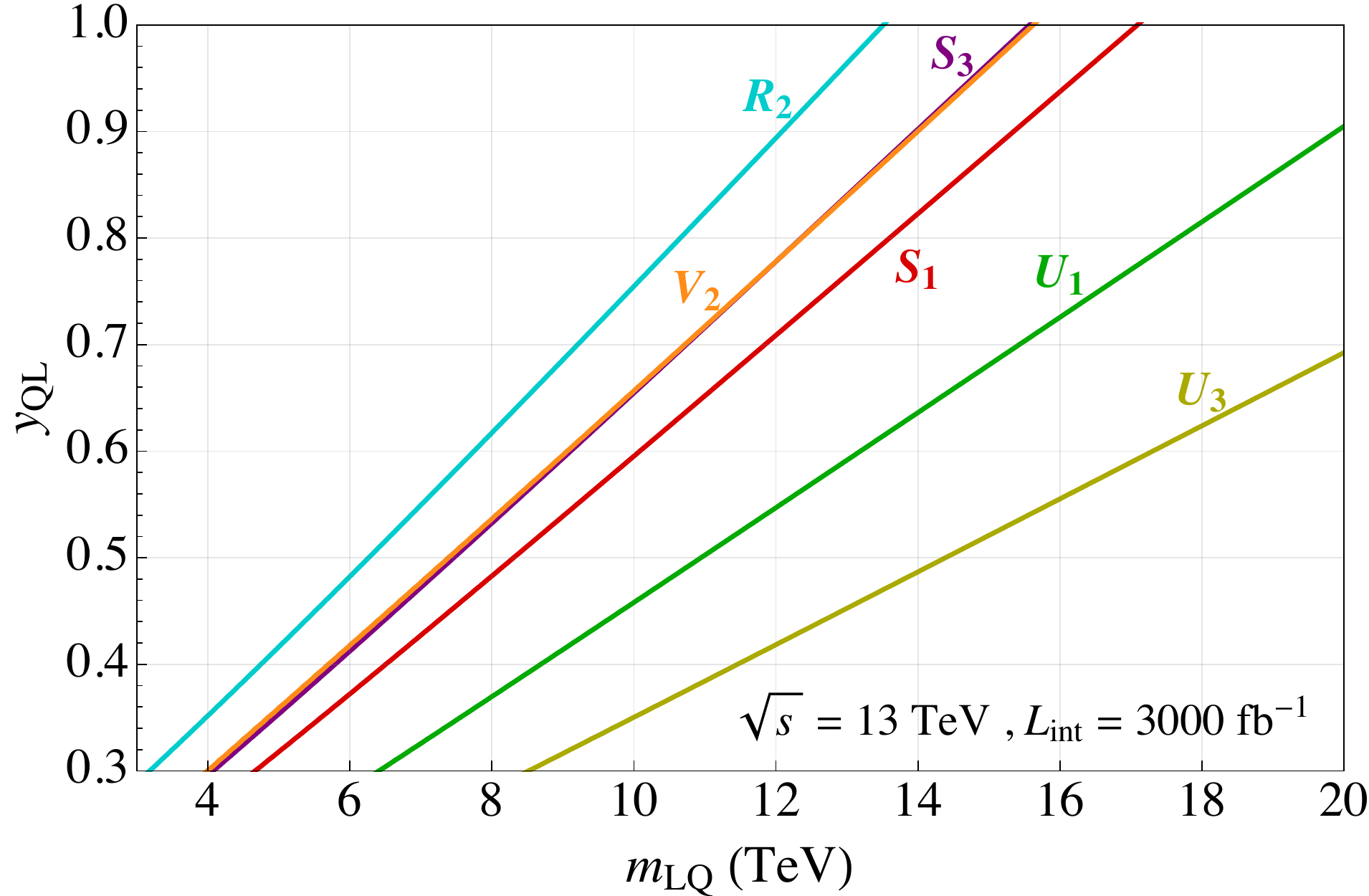}
\end{center}
\caption{
The expected 95\% \acro{C.L.} limits at the High-Luminosity (3000 fb$^{-1}$) \acro{LHC} at $\sqrt{s}$ = 13 TeV, obtained by combining the sensitivities of $p p \ra e^\pm \nu$ transverse mass and $p p \ra e^+ e^-$ invariant mass distributions.
(The limits for $S_3$ and $V_2$ being nearly identical is a coincidence.)
}
\label{fig:reaches}
\end{figure}

Having explored how leptoquarks impact Drell-Yan processes at parton level, we now proceed to determine constraints on their couplings and masses from recent \acro{LHC} Drell-Yan data. To that end, we reinterpret differential mass distributions presented by the \acro{LHC} collaborations in terms of our leptoquark parameters.
Specifically, we use spectra of 
{\sl (i)} the monolepton transverse mass $M_T \equiv [2 p^\ell_T \slashed{E}_T (1-\cos\Delta\phi)]^{1/2}$, where 
$p^\ell_T$ is the transverse momentum of the (visible) charged lepton,
$\slashed{E}_T$ is the missing transverse energy and
$\Delta\phi$ is the azimuthal opening angle between the two vectors;
and {\sl (ii)} the dilepton\footnote{In the dilepton channel there are additional experimental handles for hunting leptoquarks.
 In \cite{1610.03795}, it was shown that the scattering angle of the final state leptons was also a powerful probe, and had the further benefit of helping to determine the spin of the leptoquark. 
Due to our focus on the $q\bar{q}' \ra \ell^\pm\nu_\ell$ channel, in which the scattering angle cannot be measured due to the inability to fully reconstruct events with missing energy, we will not consider the angular signals in our analysis.
We expect, from previous work, that measurements of $d\sigma/dm_{\ell\ell}$ will give bounds on leptoquarks similar to those one would obtain from the angular distributions.} invariant mass $\mll~(= \sqrt{\hat{s}})$. 
We use measurements by \acro{ATLAS} at $\sqrt{s}$ = 13 TeV with integrated luminosity 36 fb$^{-1}$ \cite{1706.04786,1707.02424}.
We expect similar constraints from \acro{CMS} measurements \cite{1803.06292,1803.11133}, though we do not use them here since distributions are presented as events/GeV in unevenly spaced bins, which makes the recasting procedure challenging.

Because four of our leptoquark species produce amplitudes that interfere with the corresponding \acro{SM} process, we must generate event distributions in $M_T$ (for monoleptons) and $\mll$ (for dileptons) for the signal and the irreducible background together.
To that end, we employ the following procedure. 
Using the {\tt Universal FeynRules Output} files prepared by \cite{1801.07641}, the mass spectra were first generated at next-to-leading order (\acro{NLO}) in \acro{QCD} using {\tt MadGraph}\_{\tt aMC}$@${\tt NLO 2.6x} \cite{1405.0301}, 
with {\tt NN23NLO} parton density functions (PDFs) \cite{1308.0598}, setting the common renormalization and factorization scale to $\sqrt{\hat{s}}$.
As done in \cite{1709.00439}, our spectra are then rescaled by a global factor that accounts for the lepton reconstruction efficiency, so that our background spectra match the irreducible background taken from \acro{ATLAS} (and extraced via \acro{HEPData}~\cite{Maguire:2017ypu}).
The net signal and background events are obtained by adding the reducible background extracted from~\cite{1706.04786,1707.02424} to our generated events.
To quantify the effect of our signals and estimate limits on leptoquark parameters, we use the following $\chi^2$ test:
\beq
\chi^2 = \sum_i^{\rm bins} \frac{(N_{{\rm model}_i} - N_{{\rm data}_i})^2}{N_{{\rm data}_i}+\delta^2_{\rm sys}}~,
\label{eq:chisq}
\eeq
where $N_{{\rm model}_i}$ is the number of events predicted either by the \acro{SM} or by new physics and \acro{SM} together. 
The 95\% \acro{C.L.} bound is located where the difference in $\chi^2$ for signal and background models is 5.99.
Here the systematic error $\delta_{\rm sys}$, taken to be a flat 6\% as seen across multiple bins in the \acro{ATLAS} searches \cite{1706.04786,1707.02424}, is assumed uncorrelated across bins, as the co-variance matrix has not been provided.

In Fig.~\ref{fig:signalsmassproto} we show (proton-level) Drell-Yan mass distributions for each of the six leptoquark scenarios with $(\yql, \mlq$/TeV) = (0.7, 1), along with the \acro{SM} distribution. 
Importantly, the prominent interference features observed at the parton level in Fig.~\ref{fig:signalsmass} are still evident in the monolepton $M_T$ distribution and, to a lesser degree, in the dilepton $\mll$ distribution.

Turning to the data and carrying out the $\chi^2$ test of Eq.~\eqref{eq:chisq}, we interpret the observed limits on $N_{{\rm model}_i}$ as contours in the $\mlq$--$\yql$ plane. 
The results are shown in Fig.~\ref{fig:limits}.
For three of the leptoquark species ($S_1$, $S_3$, $U_1$), the monolepton $M_T$ spectrum places a stronger constraint on the leptoquark mass and coupling than the dilepton $\mll$ spectrum. For two others ($U_3$ and $V_2$), the monolepton and dilepton constraints are nearly identical in their reach, with the monolepton only slightly weaker. Only in the case of the $R_2$ leptoquark does the dilepton channel constrain the parameter space significantly better than the monolepton; this is not surprising given the much greater deviations seen in the dilepton $\mll$ spectrum as compared to the monolepton $M_T$ spectrum in Fig.~\ref{fig:signalsmassproto}. Thus, in five of the six cases studied, the monolepton channel either dominates or significantly contributes to the constraints on the allowed parameter space for that particular leptoquark (for the benchmark coupling structure chosen).

In examining the relative strength of the monolepton channel for each of the six leptoquark species, one sees two clear trends. 
First, bounds on vector leptoquarks are stronger than those on the closest corresponding scalar leptoquark, due to a relative factor of 2 that arises in the calculations of their amplitudes as well as the rapid increase in the vector leptoquark cross-sections as $\cos\theta^*\to 1$ at large $\sqrt{\hat s}$. 
Second, bounds on leptoquarks that exhibit interference with the corresponding \acro{SM} amplitude are also stronger, as one would expect; this is especially true for scalars. A similar pattern is not immediately obvious in the dilepton channel, since interference for dileptons is completely generic.

Fig.~\ref{fig:limits} also shows other relevant constraints.
Limits on the {\tt electroquarks} from \acro{LHC} pair production are taken from \cite{1706.05033}, which recast a 2.6 fb$^{-1}$ \acro{CMS} search \cite{CMS:2016imw}; these are the strongest limits to date from dedicated searches for pair-produced first-generation leptoquarks, and they appear as a vertical line since pair-production is independent of $\yql$ for $\yql \lsim 1$.
The limit on the $S_1$ {\tt electroquark} is $\mlq \geq 930$ GeV, too weak to be shown on our plot.
The 2$\sigma$ limits from \acro{APV} measurements by Wood {\em et al}. \cite{Wood:1997zq} on {\tt electroquarks} are obtained using the formulae in \cite{1603.04993}. 
We see that Drell-Yan monolepton and dilepton constraints are always stronger than \acro{APV} in constraining the leptoquark parameter space for these six leptoquarks, and are stronger than direct search/pair production bounds for leptoquark masses above $1$-$2\,$TeV and $\yql$ larger than $O(0.1$-$0.3)$.

The above conclusions are a significant update on \cite{1610.03795}, where bounds on leptoquarks were placed using only dilepton (and not monolepton) measurements, from Run 1 of the \acro{LHC}. 
In that work it was found that \acro{APV} measurements outperformed dilepton measurements for {\tt electroquarks};
our dilepton constraints are now stronger due to the higher energy and luminosity at the \acro{LHC}, which has resulted in a larger statistical sample.

We remind the reader that our dilepton limits were obtained by switching off the leptoquark coupling to right-handed fermions, {\it i.e.}\/ the coupling that generates dilepton but no monolepton signals.
We find that if we set this coupling equal to $\yql$, for $S_1$ the dilepton limits become comparable to the monolepton limits, whereas for $U_1$ they remain much weaker than monolepton limits. We also note that constraints from ``single production" processes (with signature $\ell^+\ell^-j$) may also apply, but have not been examined here since we expect them to be weaker than Drell-Yan limits, due to the energy cost of producing an on-shell leptoquark, as well as larger backgrounds and uncertainties \cite{1610.03795}.

Finally, one can check that our results reproduce the limits derived from studies of compositeness/contact interactions in the large $\mlq$ limit, which they do. However, for masses in the range being bounded here ($\mlq\lsim 5\,$TeV) we find significant differences between our analysis and a purely contact operator analysis that shift the resulting bounds on leptoquarks by a TeV or more.

\section{Prospects \& Conclusions}
\label{sec:concs}

In this paper we have shown that the spectrum of monolepton ($\ell^\pm \nu$) production at the \acro{LHC} provides stringent limits on the couplings of leptoquarks that interact with valence quarks in the proton, for a wide range of leptoquark masses not reachable by direct production searches. 
Specifically, we have analyzed the monolepton signal for the six leptoquarks. three scalars ($S_1$, $S_3$, $R_2$) and three vectors ($U_1$, $U_3$, $V_2$), that produce monoleptons. 
These limits are competitive with, and often stronger than, the ones set by the dilepton ($\ell^+\ell^-$) signals which accompany the monoleptons. 
As summarized in Fig.~\ref{fig:limits}, spectra of the monolepton transverse mass and the corresponding dilepton invariant mass already constrain leptoquark masses in the range of a few TeV, for couplings of electroweak size. Combined, these two search channels set the most stringent bounds on these six leptoquarks to date.

Because these searches are primarily statistics-limited rather than energy-limited, 
as the \acro{LHC} collects more data, the expected reach of the monolepton and dilepton searches will significantly extend the limits on the leptoquark parameter space. 
In Fig.~\ref{fig:reaches} we plot our expected reach at the High Luminosity \acro{LHC} with 3000 fb$^{-1}$ of integrated luminosity and $\sqrt{s} = 13$ TeV. 
We calculated this reach by combining the monolepton and dilepton channels, including a conservative systematic error of 15\% and neglecting all sources of reducible background.
We find that, for a coupling strength of $g$ ($\simeq 0.64$), the expected limit is between 8 TeV (for $R_2$) and 18 TeV (for $U_3$), allowing the \acro{LHC} to probe leptoquarks at masses well above its center-of-mass energy. 
At these masses, the search for leptoquarks and the study of compositeness/contact operators should coincide quite closely.

Finally, in addition to being a discovery channel, monolepton signals can also diagnose the nature of leptoquarks.
The presence of such signals would imply a leptoquark coupling to left-handed leptons, and both up and down quarks; the presence of a signal in the dilepton channel without an accompanying signal in the monoleptons would imply a leptoquark with other coupling structures. If observed, the shape of the monolepton distribution would provide strong evidence as to the identity of the underlying leptoquark.

In light of the impressive sensitivity afforded by Drell-Yan measurements, we strongly urge the \acro{LHC} collaborations to present leptoquark interpretations when analyzing Drell-Yan events. 

~

\section*{Acknowledgments}

We thank 
Joe Bramante,
Martin Schmaltz,
and
Yiming Zhong
for discussions.
This work was partially supported by the National
Science Foundation under Grant No. PHY-1520966.
\\

\appendix

\section{Leptoquark interactions in 2-component notation}
\label{app:twocomponent}

For clarity, we provide here in 2-component notation the interactions of our prototypical leptoquarks $S_1$ and $U_1$. 
In the gauge basis the interactions of $S_1$ (Eq.~\ref{eq:LagS1}) are
\begin{align}
\lambda^{ij}_{\acro{QL}}S_1 (u_{Li}\, e_{Lj} - d_{Li}\nu_{Lj}) + \lambda^{ij}_{ue} S_1 u^{\dag}_{ci}e^{\dag}_{cj} + {\rm  h.c.}
\end{align}
Similarly, for $U_1$ (Eq.~\ref{eq:LagU1}),
\begin{align}
\lambda^{ij}_{\acro{QL}}U_{1,\mu} (u^{\dag}_{Li}\bar{\sigma}^{\mu} \nu_{Lj} + d^{\dag}_{Li}\bar{\sigma}^{\mu}\, e_{Lj}) - \lambda^{ji}_{de} U_{1,\mu} e^{\dag}_{ci}\, \bar{\sigma}^{\mu}\, d_{cj} + {\rm h.c.}
\end{align}

\section{Sign of the interference}
\label{app:betas}

As discussed frequently in the paper, the sign of interference (if there is one) between the leptoquark-mediated and \acro{SM} amplitudes could be either positive or negative.
Given a leptoquark species and its coupling structure, this sign can be determined process by process, by identifying three possible sources of a relative negative sign between the amplitudes.
These sources are: 
$\beta_1$, a sign picked up when Fierzing the fermion bilinears in one of the two amplitudes so that the two operators match;
$\beta_2$, the sign of the product of the couplings in each amplitude; this could include a minus sign between the fermion currents within each operator, such as that arising from an antisymmetric $SU(2)_W$ contraction (see Eq.~\eqref{eq:LagS1}); 
$\beta_3$, the overall sign of the leptoquark propagator.

Let us apply the above to $S_1$ and $U_1$.
The sign $\beta_1$ turns out to be positive in our set-up.
In $\beta_3$, the propagator denominator, $t - \mlq^2$ or $u - \mlq^2$, is always negative, and the sign of the numerator is determined by the leptoquark spin.
The signs $\{\beta_1, \beta_2, \beta_3 \}$ for each of the Drell-Yan processes are then as follows.
In monolepton production, 
for \acro{SM}-$S_1$ interference, 
\{$+$, $-$, $+$\}, and 
for \acro{SM}-$U_1$ interference, 
\{$+$, $+$, $-$\}.
In dilepton production, 
for \acro{SM}-$S_1$ interference, we have 
\{$+$, $-$, $+$\} and 
for \acro{SM}-$U_1$ interference, 
\{$+$, $+$, $-$\}.
In all these cases we see that the overall sign is negative, as borne out by Fig.~\ref{fig:signalsmass}.

\end{document}